\documentclass[pre,twocolumn,english,superscriptaddress,floatfix]{revtex4-1}
\usepackage[T1]{fontenc}
\usepackage{amsmath}
\usepackage{amssymb}
\usepackage{graphicx}
\usepackage{esint}

\makeatletter

\newcommand{\Imm}{\mathop{\rm Im}\nolimits}

\newcommand{\Res}{\mathop{\rm Res}\nolimits}
\newcommand{\sgn}{\mathop{\rm sgn}\nolimits}
 \@ifundefined{textcolor}{}
 {%
   \definecolor{BLACK}{gray}{0}
   \definecolor{WHITE}{gray}{1}
   \definecolor{RED}{rgb}{1,0,0}
   \definecolor{GREEN}{rgb}{0,1,0}
   \definecolor{BLUE}{rgb}{0,0,1}
   \definecolor{CYAN}{cmyk}{1,0,0,0}
   \definecolor{MAGENTA}{cmyk}{0,1,0,0}
   \definecolor{YELLOW}{cmyk}{0,0,1,0}
 }

\makeatother

\usepackage{babel}
\begin{document}

\title{Perturbation theory of a superconducting $0-\pi$ impurity quantum
phase transition}

\author{M. \v{Z}onda}

\affiliation{Department of Condensed Matter Physics, Faculty of Mathematics and
Physics, Charles University in Prague, Ke Karlovu 5, CZ-12116 Praha
2, Czech Republic}

\author{V. Pokorn\'y}

\affiliation{Institute of Physics, Academy of Sciences of the Czech Republic,
Na Slovance 2, CZ-18221 Praha 8, Czech Republic}
\affiliation{Theoretical Physics III, 
Center for Electronic Correlations and Magnetism, Institute of Physics,
University of Augsburg, D-86135 Augsburg, Germany}

\author{V. Jani\v{s}}

\email{janis@fzu.cz}

\selectlanguage{english}%

\affiliation{Institute of Physics, Academy of Sciences of the Czech Republic,
Na Slovance 2, CZ-18221 Praha 8, Czech Republic}

\author{T. Novotn\'y}

\email{tno@karlov.mff.cuni.cz}

\selectlanguage{english}%

\affiliation{Department of Condensed Matter Physics, Faculty of Mathematics and
Physics, Charles University in Prague, Ke Karlovu 5, CZ-12116 Praha
2, Czech Republic}

\date{\today}

\begin{abstract}
  A single-level quantum dot with Coulomb repulsion attached to two
  superconducting leads is studied via the perturbation expansion in
  the interaction strength. We use the Nambu formalism and the
  standard many-body diagrammatic representation of the impurity Green
  functions to formulate the Matsubara self-consistent perturbation
  expansion.  We show that at zero temperature second order of the
  expansion in its \emph{spin-symmetric} version
  yields a nearly perfect agreement with the numerically exact
  calculations for the position of the $0-\pi$ phase boundary
  at which the Andreev bound states reach the Fermi energy as
  well as for the values of single-particle quantities in the $0$-phase. We
  present results for phase diagrams, level occupation,
  induced local superconducting gap, Josephson current, and energy of
  the Andreev bound states with the precision surpassing any
  (semi)analytical approaches employed thus far.
\end{abstract}
 \maketitle
%\pacs{74.50.+r, 73.21.La, 73.63.Kv, 72.15.Qm}

\section*{Introduction}Nanostructures attached to leads with specific
properties display interesting and important quantum effects at low
temperatures. Much attention, both from
experimentalists \cite{DeFranceschi10} and theorists \cite{Yeyati11},
has been paid in recent years to a quantum dot with well separated
energy levels attached to BCS superconductors.  In particular, the
behavior of the supercurrent (Josephson current) that can flow through
the impurity in equilibrium without any external voltage bias between
two superconducting leads was in the center of interest
\cite{Jarillo06,Jorgensen06,Jorgensen09}. The Josephson current
through quantum dots with tangible on-dot Coulomb repulsion can induce
a transition signalled by the sign reversal of the supercurrent
observed experimentally
\cite{vanDam06,Cleuziou06,Jorgensen07,Eichler09,Maurand12}.

This so called $0-\pi$ transition is induced by the underlying impurity
quantum phase transition (QPT) related to the crossing of  lowest many-body
eigenstates of the system from a spin-singlet ground state with positive
supercurrent ($0$-phase) to a spin-doublet state with negative supercurrent
($\pi$-phase) \cite{Matsuura77}. In single-particle spectral properties
this transition is associated with crossing of the Andreev bound
states (ABS) at the Fermi energy as has  also been experimentally observed
\cite{Pillet10,Pillet13}. Continuous vanishing of the ABS energies
at the transition is a direct consequence of crossing of  many-body
eigenstates \cite{Pillet13} and may serve as an important consistency
check of proposed theories. The latter cover by now a broad scope of techniques
ranging from numerically exact (and computationally expensive) numerical
renormalization group (NRG) \cite{Choi04,Oguri04} suitable for zero-temperature
and finite-temperature quantum Monte Carlo \cite{Siano04,Luitz10}
to (semi)analytical methods based on expansion around the atomic limit
\cite{Glazman89,Novotny05,Meng09}, mean-field theory
\cite{Rozhkov99,Yoshioka00,Vecino03,Rodero12}, 
or formalisms specialized on the strongly correlated systems such
as slave-particles \cite{Clerk00,Sellier05} and functional renormalization
group (fRG) \cite{Karrasch08}. 

However, despite of the versatility of these approaches, there still remain vast regions of the parameter space with direct experimental relevance ($\Delta\lesssim\Gamma\lesssim U$, see, e.g., Ref.~\cite{Jorgensen07}) where most of the above approaches cannot be applied and one has to
resort either to overly heavy numerical methods (NRG or QMC) or to
conceptually flawed spin-symmetry-broken mean-field approach. 
The latter approach
is not excessively elaborate and often gives quantitatively acceptable
results \cite{Rodero12}, although
 at the expense of breaking the spin symmetry of exact solution.
In particular, spin-polarized mean-field solutions even after the
symmetrization described in Ref.~\cite{Rodero12} still exhibit at
the transition unphysical discontinuities in the ABS energies \cite{Rodero12}
and in finite-temperature supercurrents \cite{Luitz12}. 

The aim of this paper is to  provide a conceptually clean and computationally inexpensive 
generic formalism for addressing the $0-\pi$ transition in that widespread regime without strong-correlations 
(i.e., without the Kondo effect). We show that a resummed perturbation theory (PT)
incorporating the second-order dynamical corrections to the
spin-symmetric mean-field (Hartree-Fock) solution yields at zero
temperature a nearly perfect description of the $0$-phase including
the position of the phase boundary in a wide parameter range outside
of strong correlations. The precision of this solution is
unprecedented by any so far employed (semi)analytical methods
including fRG. On the other hand, the solution developed from the
non-interacting limit breaks down at the phase boundary and any
perturbative description of the $\pi$-phase and, consequently, also
finite temperatures which mix $0$ and $\pi$ solutions, remains
elusive. Although the second-order PT has been applied to this problem
previously in Ref.~\cite[Sec.~V]{Vecino03} and, especially, in a
(otherwise unpublished) part of Meng's master thesis
\cite[Ch.~4]{Meng-master09}, these studies were limited to the
particle-hole symmetric case  only (in Ref.~\cite{Vecino03} in just
two limits $\Delta/\Gamma\ll1$ and $\Delta/\Gamma\gg1$) and did not
use crossing of the ABS as the definition of the boundary of the
$0$-phase. Instead, they defined the $0-\pi$ transition by equalling the approximated Kondo temperature with the
 superconducting gap, namely $\Delta = \Gamma/(1 - \partial
  \Sigma(0)/\partial \omega)$, which however holds only qualitatively.  
The
generic character of the PT method and the proper definition of the
$0-\pi$ boundary in the Green-function formalism have thus remained unnoticed.

\section*{Results} A single impurity Anderson
model is used to simulate the quantum dot with well-separated energy
levels connected to the superconducting leads in the experimental
setup \cite{Choi04,Siano04,Luitz12,Pillet13}. The Hamiltonian of the
system consisting of a single impurity with the level energy
$\varepsilon$ and local Coulomb repulsion $U$ attached to two
superconductors reads
%\begin{subequations}
\begin{equation}
\mathcal{H}=\varepsilon\sum_{\sigma=\uparrow,\downarrow}d_{\sigma}^{\dagger}d_{\sigma}^{\phantom\dagger}+Ud_{\uparrow}^{\dag}d_{\uparrow}^{\phantom\dagger}d_{\downarrow}^{\dagger}d_{\downarrow}^{\phantom\dagger}+\sum_{s=R,L}(\mathcal{H}_{{\rm lead}}^{s}+\mathcal{H}_{T}^{s})\ ,
\end{equation}
where the BCS Hamiltonian of the leads is 
\begin{equation}
\mathcal{H}_{{\rm
    lead}}^{s}=\sum_{\mathbf{k}\sigma}\epsilon(\mathbf{k})c_{s\mathbf{k}\sigma}^{\dagger}c_{s
   \mathbf{k}\sigma}^{\phantom\dagger}-\Delta_{s}\sum_{\mathbf{k}}(e^{i\Phi_{s}}c_{s\mathbf{k}\uparrow}^{\dagger}c_{s\mathbf{-k}\downarrow}^{\dagger}+\textrm{H.c.})\ ,
\end{equation}
with $s=L,R$ denoting the left/right lead, respectively. Finally,
the hybridization term between the impurity and the contacts is given
by 
\begin{equation}
\mathcal{H}_{T}^{s}=-t_{s}\sum_{\mathbf{k}\sigma}(c_{s\mathbf{k}\sigma}^{\dagger}d_{\sigma}^{\phantom\dagger}+\textrm{H.c.})\ .
\end{equation}
%\end{subequations}

The individual degrees of freedom of the leads are unimportant for the
studied problem and are generally integrated out, leaving us with only
the active variables and functions on the impurity. Due to the
proximity effect there are locally induced superconducting
correlations on the impurity and the most direct way to handle them is
via the Nambu spinor representation of the local fermionic operators
in which the one-electron impurity (imaginary time/Matsubara) Green
function (GF) is a $2\times2$ matrix
%\begin{multline}
%\widehat{G}_{\sigma}(\tau-\tau')\equiv\begin{pmatrix}G_{\sigma}(\tau-\tau')\ , & \mathcal{G}_{-\sigma}(\tau-\tau')\\
%\bar{\mathcal{G}}_{\sigma}(\tau-\tau')\ , & \bar{G}_{-\sigma}(\tau-\tau')
%\end{pmatrix}
%=\begin{pmatrix}
%    \quad\includegraphics[width=0.22\textwidth]{gf.eps}\quad
% \end{pmatrix}\\ 
%=-\begin{pmatrix}\langle\mathbb{T}[d_{\sigma}^{\phantom\dagger}(\tau)d_{\sigma}^{\dag}(\tau')]\rangle\ , & \langle\mathbb{T}[d_{\sigma}^{\phantom\dagger}(\tau)d_{-\sigma}^{\phantom\dagger}(\tau')]\rangle\\[0.3em]
%\langle\mathbb{T}[d_{-\sigma}^{\dag}(\tau)d_{\sigma}^{\dag}(\tau')]\rangle\ , & \langle\mathbb{T}[d_{-\sigma}^{\dag}(\tau)d_{-\sigma}^{\phantom\dagger}(\tau')]\rangle
%\end{pmatrix}, \label{eq:GFdef}
%\end{multline}
%\begin{multline}
\begin{equation}
\begin{split}
\widehat{G}_{\sigma}(\tau-\tau')&\equiv\begin{pmatrix}G_{\sigma}(\tau-\tau')\ , & \mathcal{G}_{-\sigma}(\tau-\tau')\\
\bar{\mathcal{G}}_{\sigma}(\tau-\tau')\ , & \bar{G}_{-\sigma}(\tau-\tau')
\end{pmatrix}\\
&=-\begin{pmatrix}\langle\mathbb{T}[d_{\sigma}^{\phantom\dagger}(\tau)d_{\sigma}^{\dag}(\tau')]\rangle\ , & \langle\mathbb{T}[d_{\sigma}^{\phantom\dagger}(\tau)d_{-\sigma}^{\phantom\dagger}(\tau')]\rangle\\[0.3em]
\langle\mathbb{T}[d_{-\sigma}^{\dag}(\tau)d_{\sigma}^{\dag}(\tau')]\rangle\ , & \langle\mathbb{T}[d_{-\sigma}^{\dag}(\tau)d_{-\sigma}^{\phantom\dagger}(\tau')]\rangle
\end{pmatrix}, \label{eq:GFdef}
\end{split}
\end{equation}

%\end{multline}
where the bar denotes the hole function.

The impurity GF can be exactly found for an impurity without onsite
interaction ($U=0$) by method analogous to Appendix A of Ref.~\cite{Novotny05}.
When assuming identical left and right superconducting gaps $\Delta_{L}=\Delta_{R}\equiv\Delta$
as well as tunnel couplings $t_{L}=t_{R}\equiv t$ it can be written in terms
of Matsubara frequencies $\omega_{n}\equiv(2n+1)\pi/\beta$ as ($e=\hbar=1$
throughout the paper; we also skip the spin index as we only consider
spin-symmetric solutions) 
\begin{equation}
\widehat{G}_{0}(i\omega_{n})=\begin{pmatrix}i\omega_{n}[1+s(i\omega_{n})]-\varepsilon\
  , & \Delta_{\Phi} s(i\omega_{n})\\[0.3em]
\Delta_{\Phi} s(i\omega_{n})\ , & i\omega_{n}[1+ s(i\omega_{n})]+\varepsilon
\end{pmatrix}^{-1},\label{eq:D0}
\end{equation}
where $s(i\omega_{n})=\frac{\Gamma}{\sqrt{\Delta^{2}+\omega_{n}^{2}}}$
is the hybridization self-energy due to the coupling of the impurity
to the superconducting leads. We have denoted by 
$\Gamma=2\pi t^{2}\rho_{0}$ the normal-state tunnel coupling
magnitude ($\rho_{0}$ being the normal-state density of states of
lead electrons at the Fermi energy) and $\Delta_{\Phi}\equiv\Delta\cos(\Phi/2)$ 
with $\Phi=\Phi_{L}-\Phi_{R}$ being 
the difference between the phases of the left and right superconducting
leads.

The impact of the Coulomb repulsion $U>0$ on the Green function
is included in the interaction self-energy matrix  
$\widehat{\Sigma}(i\omega_{n})\equiv\left(\begin{smallmatrix}\Sigma(i\omega_{n}),
    & \mathcal{S}(i\omega_{n}) \\
\mathcal{\bar{S}}(i\omega_{n}), & \bar{\Sigma}(i\omega_{n})
\end{smallmatrix}\right)$,
so that the full propagator in the
spin-symmetric situation is determined 
by the Dyson equation $\widehat{G}^{-1}(i\omega_{n})=\widehat{G}_{0}^{-1}(i\omega_{n})-\widehat{\Sigma}(i\omega_{n})$.
The symmetry relations for the Green function equation~\eqref{eq:GFdef} reformulated
in the Matsubara representation as
$\bar{G}_{\sigma}(i\omega_{n})=-G_{\sigma}(-i\omega_{n})$ 
and $\mathcal{\bar{G}}_{\sigma}(i\omega_{n})=\mathcal{G}_{\sigma}(-i\omega_{n})$
imply the same for the self-energies, i.e.~$\bar{\Sigma}_{\sigma}(i\omega_{n})=-\Sigma_{\sigma}(- i\omega_{n})$ and
$\bar{S}_{\sigma}(i\omega_{n})=S_{\sigma}(- i\omega_{n})$. 
Therefore, the Green function explicitly reads 
\begin{equation}
\begin{split}
&\widehat{G}(i\omega_{n})=-\frac{1}{D(i\omega_{n})}\times\\
&\begin{pmatrix}i\omega_{n}[1+s(i\omega_{n})]+\varepsilon+\Sigma(-i\omega_{n}),
  & -\Delta_{\Phi} s(i\omega_{n}) +\mathcal{S}(i\omega_{n})\\[0.3em]
-\Delta_{\Phi} s(i\omega_{n})+\mathcal{S}(-i\omega_{n}), &
i\omega_{n}[1+ s(i\omega_{n})]-\varepsilon-\Sigma(i\omega_{n})
\end{pmatrix}.
\end{split}\label{eq:GF}
\end{equation}

The negative determinant of the inverse Green function
$D(i\omega_{n})\equiv-\det[\widehat{G}^{-1}(i\omega_{n})]=\omega_{n}^{2}\left[1+s(i\omega_{n})\right]^{2}+\left[\varepsilon+\Sigma(i\omega_{n})\right]
\left[\varepsilon+\Sigma(-i\omega_{n})\right]+\left[\Delta_{\Phi}
  s(i\omega_{n})-\mathcal{S}(i\omega_{n})\right]
\left[\Delta_{\Phi}s(i\omega_{n})-\mathcal{S}(-i\omega_{n})\right]$  
determines via its zeros the existence and positions of the ABS. This
determinant is real within the gap and can go through zero $D(\omega_{0})=0$
determining the (real) in-gap energies $\pm\omega_{0}$ of the ABS symmetrically
placed around the Fermi energy (center of the gap). The ABS are important
for transport of the Cooper pairs through the quantum dot and
usually provide the  dominant contribution to the dissipation-less Josephson
current $J$ through the impurity, which can be evaluated at zero
temperature by an integral of the anomalous Green function
(see the Methods section) 
\begin{equation}
\begin{split}\frac{J}{4\Delta} &
  =-\int_{-\infty}^{\infty}\frac{d\omega_{n}}{2\pi}\
  \Im\left[\mathcal{G}(i\omega_{n}) s(i\omega_{n})e^{-i\frac{\Phi}{2}}\right]\\
 & =-\Gamma\sin\frac{\Phi}{2}\left[\frac{\mathrm{Res}(\mathcal{G};-\omega_{0})}{\sqrt{\Delta^{2}-\omega_{0}^{2}}}+\int_{-\infty}^{-\Delta}\frac{d\omega}{\pi}\frac{\Re\mathcal{G}(\omega)}{\sqrt{\omega^{2}-\Delta^{2}}}\right].
\end{split}
\label{eq:JC}
\end{equation}
While the first line uses the thermal representation via Matsubara
frequencies the second one is the analytic continuation to the real
frequencies (spectral representation) which allows us to distinguish
the direct supercurrent through the lower ABS (corresponding to the
residue of the anomalous impurity Green function at the negative ABS
frequency) from the tunneling current between the continuum band states below the SC gap.

\subsection*{Spin-symmetric Hartee-Fock approximation} 
As the exact expression
for this model's self-energy is unknown we resort to the standard
Matsubara perturbation theory summing one-particle irreducible diagrams
for the self-energy.

 \begin{figure}
    \includegraphics[width=0.88\columnwidth]{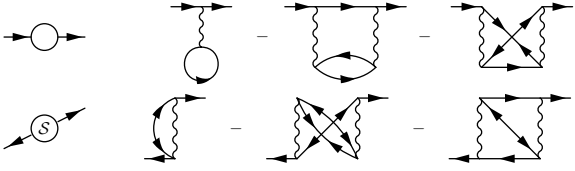}
    \caption{Diagrammatic representation of the first two orders of
      the perturbation expansion in the Coulomb interaction for the
      normal (top) and anomalous (bottom) parts of the self-energy. The wavy line
      represents the Coulomb interaction and the lines with single
      (double) arrow represent the normal (anomalous) propagators
      according to equation~\eqref{eq:GFdef}. \label{Fig1sup}}
\end{figure}

The simplest diagrams are the first-order Hartree-Fock contributions
represented by the first diagrams on the r.h.s. of equations in
Fig.~\ref{Fig1sup}. Their mathematical equivalents read
\begin{equation}
  \label{EQ:HFse}
  \Sigma^{HF}=\frac{U}{\beta}\sum_{n\in\mathbb{Z}}G(i\omega_n)\quad\mathrm{and}\quad
  \mathcal{S}^{HF}=\frac{U}{\beta}\sum_{n\in\mathbb{Z}}\mathcal{G}(i\omega_n).
\end{equation}

\begin{figure}
\includegraphics[width=\columnwidth]{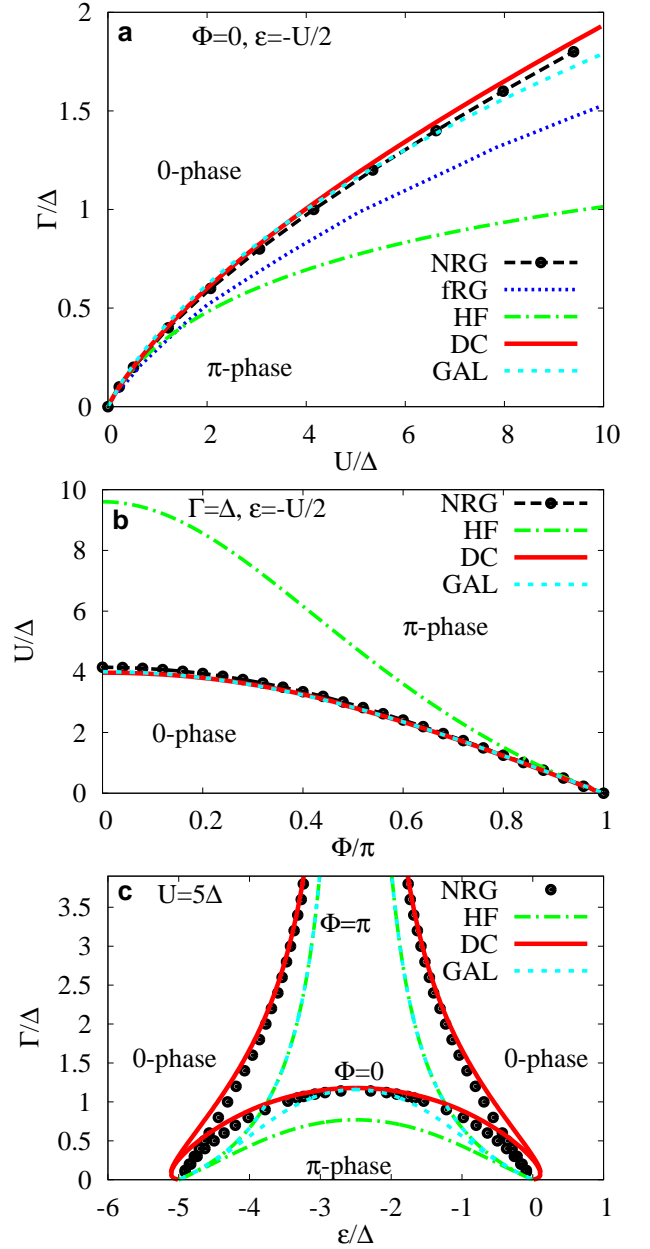}
\protect\caption{Phase diagrams in the $\Gamma-U$ (\textbf{a}), $U-\Phi$ (\textbf{b}), and $\Gamma-\epsilon$ (\textbf{c})
parameter planes. We compare the phase boundaries calculated by numerically
exact NRG with various analytical approximations: fRG (only in panel
\textbf{a}; data taken graphically from Fig.~2 of Ref.~\cite{Karrasch08}),
spin-symmetric HF, the second-order PT/dynamical corrections (DC),
and generalized atomic limit approximation (GAL) $U^{2}/(1+\Gamma/\Delta){}^{2}=(2\varepsilon+U)^{2}+4\Gamma^{2}\cos^{2}(\Phi/2)$.
\label{fig:Phase-diagrams}}
\end{figure}

The HF approximation leads just to a static, frequency-independent
mean-field self-energy neglecting any dynamical correlations caused
by particle interaction. Despite of this simplicity and contrary to
the common belief, this approximation yields \emph{without any symmetry breaking}
the $0-\pi$ quantum phase transition and we thus use it as
a convenient and sufficiently simple demonstration of the generic
features of the perturbation expansion.  The Hartree-Fock approximation
consists of two self-consistent non-linear equations that can
be reformulated in terms of auxiliary quantities
$E_{d}=\varepsilon+U\left\langle  d_{\sigma}^{\dagger}d_{\sigma}\right\rangle $ 
(mean-field energy of the level) and $\delta\equiv\Gamma\cos(\Phi/2)+\Delta_{d}$
(related to the locally induced gap $\Delta_{d}\equiv-U\left\langle
  d_{\downarrow}d_{\uparrow}\right\rangle $). They read
\begin{equation*}
%\begin{split}
E_{d} =\varepsilon+\frac{U}{2}-\frac{U}{\beta}\sum_{n\in\mathbb{Z}}  \frac{E_{d}}{D^{\mathrm{HF}}(i\omega_{n})}\,,
\end{equation*}
\begin{equation}
\delta =\Gamma\cos\frac{\Phi}{2}-\frac{U}{\beta}\sum_{n\in\mathbb{Z}}
\frac{\delta-\Gamma\cos\frac{\Phi}{2}\left(1-\frac{\Delta}{\sqrt{\omega_{n}^{2}
        +\Delta^{2}}}\right)}{D^{\mathrm{HF}}(i\omega_{n})} \ .
%\end{split}
\label{eq:HF-self-consistent}
\end{equation}
Since we are primarily interested in the zero-temperature QPT where
the energies of the  ABS approach zero $\omega_{0}\to0$, we can 
approximate 
the denominators  in the integrals by their low-frequency asymptotics
$D^{\mathrm{HF}}(i\omega\to0) \approx
E_{d}^{2}+\delta^{2}+(1+\Gamma/\Delta)^{2}\omega^{2}$, 
which implies $\omega_{0}\approx\sqrt{E_{d}^{2}+\delta^{2}}/(1+\Gamma/\Delta)$.
Near the quantum critical point we then obtain  
\begin{equation}
\begin{split}E_{d}\left[1+\frac{U}{2\omega_{0}\left(1 +
        \frac{\Gamma}{\Delta}\right){}^{2}}\right]   & =\varepsilon+\frac{U}{2}\,,\\ 
\delta\left[1+\frac{U}{2\omega_{0}\left(1
      +\frac{\Gamma}{\Delta}\right){}^{2}}\right] &
=\Gamma\cos\frac{\Phi}{2}
\left[1+\frac{U}{\Delta}\mathcal{I}\left(\frac{\Gamma}{\Delta},\Phi\right)\right]\,, 
\end{split}
\end{equation}
with the band contribution expressed via the function\\
$\mathcal{I}(x,\Phi)=\int_{0}^{\infty}\tfrac{dt}{2\pi}\tfrac{\cosh^{2}t}{\cosh^{2}(t/2)(x + \cosh t)^{2}+x^{2}\cos^{2}(\Phi/2)\sinh^{2}(t/2)}$.
Re-parametrizing \newline $E_{d}=(1+\Gamma/\Delta)\omega_{0}\cos\psi$
and $\delta=(1+\Gamma/\Delta)\omega_{0}\sin\psi$ we arrive
at 
\begin{equation}
\begin{split}\left[\frac{U}{2\left(1+\frac{\Gamma}{\Delta}\right)^{2}}
    +\omega_{0}\right]\cos\psi &
  =\frac{\varepsilon+\frac{U}{2}}{1+\frac{\Gamma}{\Delta}}\,,\\ 
\left[\frac{U}{2\left(1+\frac{\Gamma}{\Delta}\right)^{2}}+\omega_{0}\right]\sin\psi
&  =\frac{\Gamma\cos\frac{\Phi}{2}}{1+\frac{\Gamma}{\Delta}} \left[1+
  \frac{U}{\Delta}\mathcal{I}\left(\frac{\Gamma}{\Delta},\Phi\right)\right]\,. 
\end{split}
\label{eq:around-QPT}
\end{equation}

\begin{figure}
\includegraphics[width=\columnwidth]{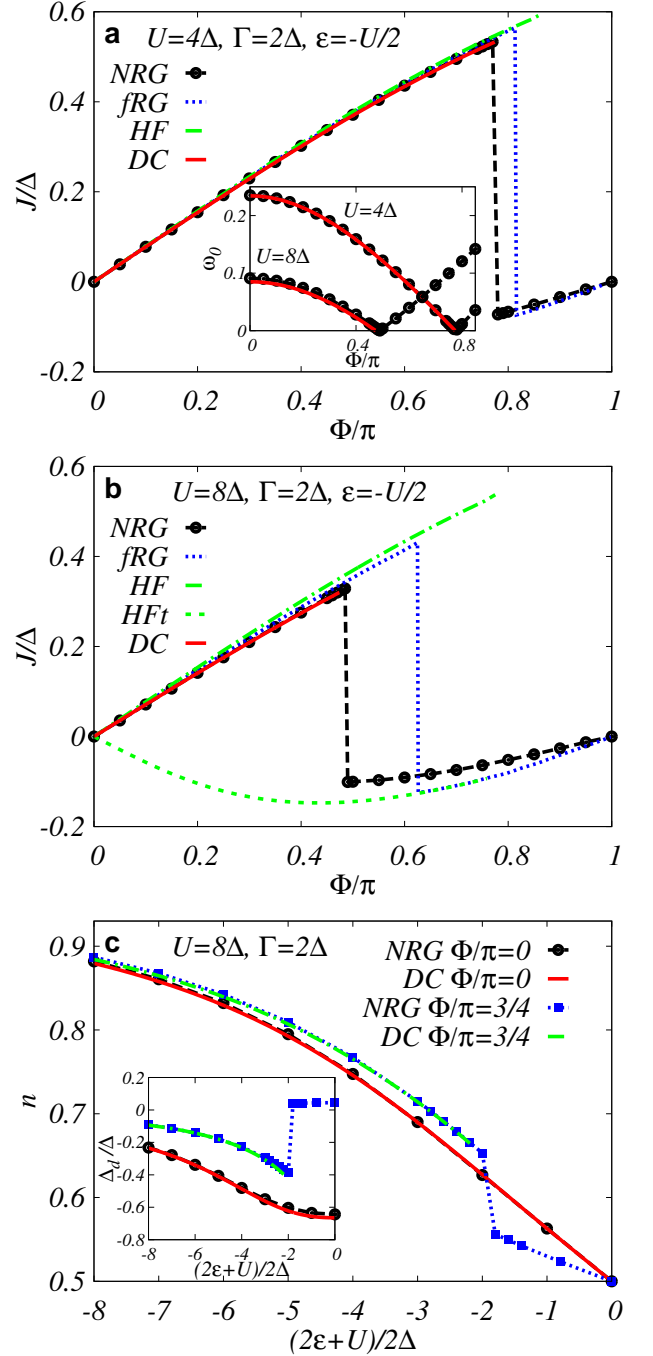}
\protect\caption{Comparing various methods of calculation of one-particle quantities.
Panels \textbf{a} and \textbf{b} show supercurrent at half-filling
as a function of the phase difference $\Phi$ for $U=4\Delta$ (\textbf{a})
and $U=8\Delta$ (\textbf{b}) calculated by numerically exact NRG,
and analytically approximative fRG, spin-symmetric HF and, finally,
the second-order PT/dynamical corrections (DC) showing a nearly perfect
agreement with NRG (unlike the other two methods). Inset in panel
\textbf{a} depicts the ABS energies $\omega_{0}$ as functions of
$\Phi$ for the two values of the Coulomb interaction $U$. The green dashed line in panel \textbf{b} represents the HF tunneling current component.
In panel \textbf{c} the occupation number $n=\left\langle d_{\sigma}^{\dagger}d_{\sigma}\right\rangle $
and locally induced SC gap $\Delta_{d}\equiv-U\left\langle d_{\downarrow}d_{\uparrow}\right\rangle $
(inset) are plotted as functions of the level energy for two values
of the phase difference $\Phi=0$ (with no phase transition) and $\Phi=\pi$
(exhibiting phase transition). fRG data in panels \textbf{a}, \textbf{b}
were graphically taken from Fig.~4b of Ref.~\cite{Karrasch08}.
\label{fig:one-particle-quantities}}
\end{figure}

At the QPT characterized by $\omega_{0}=0$ the solubility
condition ($\cos^{2}\psi+\sin^{2}\psi=1$) gives us the equation
for the HF phase boundary 
\begin{equation}
\left[\frac{U}{2\left(1+\frac{\Gamma}{\Delta}\right)}\right]^{2}=\left[\varepsilon
  +\frac{U}{2}\right]^{2}+\Gamma^{2}\cos^{2}\frac{\Phi}{2}
\left[1+\frac{U}{\Delta}\mathcal{I}
  \left(\frac{\Gamma}{\Delta},\Phi\right)\right]^{2} \label{eq:phase-boundary}
\end{equation}
that generalizes the corresponding well-known expression in the atomic
limit $\Delta\to\infty$ \cite{Bauer07,Karrasch08}.
This HF
phase boundary plotted in Fig.~\ref{fig:Phase-diagrams} is not particularly
precise, however, it yields qualitatively reasonable results. Moreover,
we have noticed that when the band contribution $\mathcal{I}$ in
equation~\eqref{eq:phase-boundary} is omitted one gets a surprisingly
good and extremely simple approximation for the boundary, that we
call here  the \emph{generalized atomic limit} (GAL), lying for half-filling ($\epsilon=-U/2$) typically very
close to the numerically exact results by NRG, see Fig.~\ref{fig:Phase-diagrams}a-b.
Obviously, the HF approximation heavily overestimates the contribution
from the band continuum. 

Eqs.~\eqref{eq:around-QPT} may be used also around the QPT, when
$\omega_{0}$ is small (and unknown). We can see that
$\omega_{0}$ is positive on one side of the boundary while it
is negative on the other side. Since $ \omega_{0}>0$ by
construction, we must conclude that the solution with negative
$\omega_{0}$, that we identify with the $\pi$-phase
region, is unphysical. We cannot go beyond the phase boundary from the $0$-phase to
the $\pi$-phase within this perturbative approach based on the assumption of a
nondegenerate ground state.

\subsection*{Dynamical corrections}
The qualitative predictions of the
HF approximation can be significantly improved by including dynamical
corrections into the self-energy, which come from the second order
of the perturbation expansion
represented by the second and third diagrams on the r.h.s. of equations in
Fig.~\ref{Fig1sup}. The two diagrams originate in two different types of
intermediate propagation consisting of either normal or anomalous
propagator. The mathematical equivalents for the second-order
contributions read
%\begin{subequations}
\begin{align}
  \Sigma^{(2)}(i\omega_n)&=-\frac{U^2}{\beta}\sum_{m\in\mathbb{Z}} 
  G(i\omega_n+i\nu_m)\chi(i\nu_m) \label{EQ:2ndse1}\\
 \intertext{and}
  \mathcal{S}^{(2)}(i\omega_n)&=-\frac{U^2}{\beta}\sum_{m\in\mathbb{Z}} 
  \mathcal{G}(i\omega_n+i\nu_m)\chi(i\nu_m)
\end{align}
where
\begin{gather}  
  \chi(i\nu_m)= \frac{1}{\beta}\sum_{n\in\mathbb{Z}}
  \left[G(i\omega_n)G(i\omega_n+i\nu_m)
    +\mathcal{G}(i\omega_n)\mathcal{G}(i\omega_n+i\nu_m)\right]\label{EQ:2ndse3}
\end{gather}
%\end{subequations}
is the two-particle bubble consisting of the normal and anomalous
parts and $\nu_m=2\pi m/\beta$ is the $m$-th bosonic Matsubara
frequency.

These first two orders of the perturbation expansion are well
controllable on the one-particle level.  The higher contributions to
the self-energy become more complex and their classification more
complicated.  For a general discussion of this problem see Ref.~\cite{Janis14}.
 
The second order self-energy correction together with the first-order (in $U$) HF
counterparts are inserted into the equation for the Green function,
equation~\eqref{eq:GF}. We obtain a self-consistent nonlinear functional
equation for the whole Green function as a function of frequency. This
equation is solved numerically at zero temperature. We noticed,
however, that nearly identical results are obtained by computationally
less elaborate method which evaluates the dynamical
self-energies  by using just a fully converged HF
solution as the input GF.  The convolutions in
the second-order self-energies are thus evaluated just once at the
beginning of the procedure and consequently used as fixed inputs into
the self-consistent procedure iterating the Green function through the
HF self-energy. It should be stressed that while the second-order
contribution may be simplified in this way, the full self-consistency
loop between the GF and the HF self-energy is mandatory --- any
compromises there lead to even qualitatively wrong results.

\section*{Discussion}
We have carried out the above mentioned procedure both in the Matsubara
formalism as well as in the spectral representation (performing the analytic
continuation described in Methods) with identical results. We have found that the 
$0$-phase smoothly develops from the noninteracting limit $U=0$
and terminates at the $0-\pi$ phase boundary beyond which there exists no regular self-consistent
solution for the GF. In the spectral representation
this is associated with the energy of ABS $\omega_{0}$ reaching zero.
The results for the phase boundaries, shown in Fig.~\ref{fig:Phase-diagrams},
and one-particle quantities in the $0$-phase in Fig.~\ref{fig:one-particle-quantities}
exhibit unprecedented precision of the dynamical corrections approximation
which gives numerical results nearly identical to the numerically
exact NRG data produced by the ``NRG Ljubljana'' open source code
\cite{NRGLjubljana,Zitko09} for all studied parameter sets as well as all
physical quantities. Surprisingly, it outperforms in the regime of not-so-weak interaction even the fRG method
designed for the strong correlations
(see the $U$-axis scale in Fig.~\ref{fig:Phase-diagrams}a).
This is likely due to the static-vertex implementation of the fRG
in Ref.~\cite{Karrasch08}. The limitations of the static-vertex approximation have been discussed before 
(see Ref.~\cite{Karrasch-PhD10}, Sec.~9.4.6), nevertheless it is currently the only
one technically viable for fRG. 
On the other hand our dynamical corrections
properly include the frequency dependence of the correlation effects
(even if just perturbatively) which probably explains their superiority
over the fRG in the description of $0$-phase quantities as well as
the phase boundary.
In this context, we would also like to point out an interesting observation we have made. 
In Fig.~\ref{fig:one-particle-quantities}b we plot (by the green dashed line) the tunnelling part of the supercurrent (the second term in the lower equation~\eqref{eq:JC}) for the HF solution and see that it coincides in the overlapping range of parameters with the full supercurrent solution of the fRG in the $\pi$-phase. Although plotted for clarity just in Fig.~\ref{fig:one-particle-quantities}b this observation holds for all $J-\Phi$ characteristics taken graphically from Ref.~\cite{Karrasch08}.
Since our HF solution breaks down at the phase boundary we cannot extrapolate beyond it, nevertheless there is obviously some subtle correspondence between
the spin-symmetric HF solution and the $\pi$-phase solution of the fRG.                  

To conclude, we have presented a systematic perturbative
expansion for the $0-\pi$ transition in the superconducting Anderson
model and found out that its second order yields at zero temperature
excellent results for the phase boundary and quantities in the
$0$-phase such as locally induced superconducting gap or supercurrent
surpassing any (semi)analytical methods employed to this model so
far.  Although demonstrated here explicitly just for the symmetric case
 $\Gamma_{L}=\Gamma_{R}$ for simplicity, the method produces equally
 good results also in the general case. Moreover, we have also verified numerically
that the formalism is gauge-invariant, i.e., physical quantities depend on the phase difference $\Phi_{L}-\Phi_{R}$
 only and conserves current, i.e., supercurrents calculated at left/right
 junctions are identical. Furthermore, the full second-order PT is  thermodynamically consistent (unlike, e.g., fRG \cite{Karrasch-PhD10}).

The method cannot be, however, continued to the $\pi$-phase
without modifications taking into account the degeneracy of the
doublet ground state. Moreover, we have observed that the Matsubara formalism
at finite temperatures does not detect any sharp phase boundary 
found at $T=0$.  To our best knowledge there is presently no
(semi)analytical method that would conceptually correctly and quantitatively reasonably describe the
$\pi$-phase. The spin-polarized HF suffers from the
discontinuity problems mentioned in the Introduction while the fRG solution 
returns $\varepsilon$- and $U$-independent quantities in the $\pi$-phase \cite{Karrasch08} apparently closely related with the simplest spin-symmetric Hartree-Fock approximation as discussed above, which is clearly not sufficient.  The construction of an analytical theory of the $\pi$-phase thus remains an open
challenge for future study.

%------------------------------------------------------------------------------------------------------------
\section*{Methods}
The necessary information for the study of the crossing of ABS as well 
as for obtaining the particular components of total current can not be obtained directly from the
expressions in Matsubara frequencies.      
To access it we analytically continued the expressions to the real-frequency domain.

\begin{figure}[ht]
  \begin{center}
    \includegraphics[width=0.8\columnwidth]{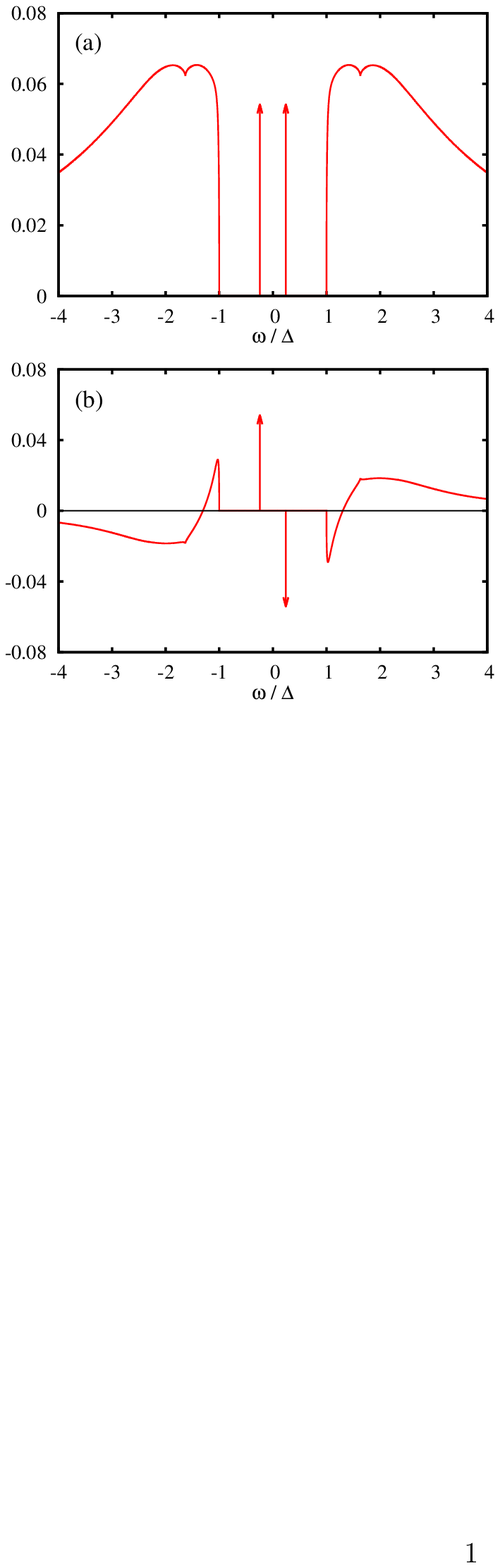}
    \caption{Normal ($-\Imm G/\pi$, upper panel (a)) and anomalous ($-\Imm
      \mathcal{G}/\pi$, lower panel (b)) spectral density for $U=4\Delta$,
      $\Gamma=2\Delta$, $\Phi=\pi/2$ and $\varepsilon=-U/2$
      (half-filling) calculated using the dynamical corrections from
      the second-order of the perturbation expansion. The heights of
      the arrows marking the Andreev bound states represent their
      residues.\label{Fig2}}
  \end{center}
\end{figure}

The inverse Green function~\eqref{eq:GFdef} can be represented as
\begin{equation}
  \hat{G}^{-1}(z)=
  \begin{pmatrix}
    z[1+s(z)]-\varepsilon-\Sigma(z) & \Delta_\Phi s(z)-\mathcal{S}(z) \\[0.3em]
    \Delta_\Phi s(z)-\mathcal{S}(-z) &
    z[1+s(z)]+\varepsilon+\Sigma(-z)
  \end{pmatrix}
\end{equation}
where
\begin{equation}
  s(z)=-\frac{i\Gamma}{\zeta}\sgn(\Imm z)
\end{equation}
is a dynamical renormalization of the impurity energy level due to the
hybridization to the superconducting leads.  We introduced a
renormalized complex energy $\zeta=\xi+i\eta$ related to $z=\omega+iy$
via $\zeta^2=z^2-\Delta^2$.  The following convention for complex
square root is used:
\begin{equation}
  \xi\eta=\omega y, \qquad \sgn\xi = \sgn \omega, \qquad \sgn\eta = \sgn y,
\end{equation}
so that $\zeta=z$ for $\Delta=0$. The renormalized energy $\zeta$
along the real axis is then real outside the energy gap and imaginary
within it. Accordingly to this definition the function $s(z)$ is
imaginary outside the energy gap and real within it,
\begin{equation}
  \begin{array}{ll}
    s(\omega\pm i0)=\pm\dfrac{i\Gamma\sgn(\omega)}{\sqrt{\omega^2-\Delta^2}}\quad & \mathrm{for}\quad|\omega|>\Delta,\\
    s(\omega\pm i0)=\dfrac{\Gamma}{\sqrt{\Delta^2-\omega^2}}\quad & \mathrm{for}\quad|\omega|<\Delta.
  \end{array}
\end{equation}
This definition allows for a straightforward analytic continuation
of the Matsubara Green function to real frequencies.  An illustrating
example of the normal and anomalous spectral functions is plotted in
Fig.~\ref{Fig2}.  The Green function has a gap around the Fermi energy
from $-\Delta$ to $\Delta$ and two poles at $\pm\omega_0$,
$|\omega_0|<\Delta$. The positions of these poles are given by zeroes
of the determinant, $\det[\hat{G}^{-1}(\omega_0)]=0$.  Since the
function $s(\omega)$ has a square-root singularity at gap edges, the
gap is fixed and does not depend on interaction strength.

Calculating the self-energy from diagrammatic expansion calls for the
analytic continuation of sums over Matsubara frequencies. The sum of a
one-particle function $F$ over fermionic Matsubara frequencies can be
rewritten in the spectral representation as \cite{Mahan00}
\begin{equation}
  \begin{split}
  \frac{1}{\beta}\sum_{n\in\mathbb{Z}} F(i\omega_n)\rightarrow
  &-\frac{1}{\pi}\left[\int_{-\infty}^{-\Delta}\!\!+\int_{\Delta}^{\infty}\right]
  d\omega f(\omega)\Imm F(\omega+i0)\\&+\sum_if(\omega_i)\Res(F,\omega_i),
  \end{split}
\end{equation}
where $\omega_i$ are the isolated poles within the gap and $f(\omega)$
is the Fermi-Dirac function. This formula can be used directly to
calculate the static Hartree-Fock self-energies~\eqref{EQ:HFse} and
the Josephson current~\eqref{eq:JC}.

Similar approach can be utilized to calculate the two-particle bubbles
and the second-order dynamic corrections,~Eqs.~\eqref{EQ:2ndse1}-\eqref{EQ:2ndse3}. For
the sake of simplicity we resort to zero temperature. Choosing a
correct contour in the upper complex half-plane we arrive at an
expression for the normal part of the bubble,
\begin{equation}
  \begin{split}
  \chi_n(\omega^+)=&-\frac{1}{\pi}\int_{-\infty}^{-\Delta}\mkern-18mu
  dx\Imm G(x^+)\left[G(x+\omega^+)+G(x-\omega^+)\right]\\
  &+\Res(G,-\omega_0)[G(-\omega_0+\omega)+G(-\omega_0-\omega)]
  \end{split}
\end{equation}
and analogously for the anomalous part $\chi_a$. We have abbreviated
$\omega^+=\omega+i0$.  The resulting bubble has an extended gap from
$-\Delta-\omega_0$ to $\Delta+\omega_0$. The contributions from the
isolated states at $\pm2\omega_0$ from the normal and anomalous parts
exactly cancel out each other, so there are no gap states in the full
bubble $\chi=\chi_n+\chi_a$. Taking this into consideration we arrive
at a formula for the normal self-energy,
\begin{equation}
 \begin{split}
  \Sigma^{(2)}(\omega^+) & =\frac{U^2}{\pi}\int_{-\infty}^{-\Delta}\mkern-18mu 
    dx\Imm G(x^+)\chi(x-\omega^+)\\
    &+\frac{U^2}{\pi}\int_{-\infty}^{-\Delta-\omega_0}\mkern-44mu
    dx\Imm \chi(x^+)G(x+\omega^+)\\
    &-U^2\Res(G,-\omega_0)\chi(-\omega_0-\omega)
    \end{split}
\end{equation}
and similarly for $\mathcal{S}^{(2)}$.  Integrals of this kind can be
evaluated numerically using fast Fourier transform algorithms which
makes the calculation simple and efficient.
%---------------------------------------------------------------------------------------------

\textbf{Acknowledgments} V.J. and V.P. thank D. Shapiro for many fruitful
discussions during his stay at the Institute of Physics, AS CR. M.\v{Z}.
thanks R.~\v{Z}itko for advice concerning the ``NRG Ljubljana'' code.
Research on this problem was supported in part by Grant P204-11-J042
of the Czech Science Foundation.

\bibliography{JosephsonAbb}
\end{document}